\documentclass{aa}  

\usepackage{rotating}           
\usepackage{tabularx}           
\usepackage{color}              
\usepackage{comment}            
\usepackage{caption}            
\usepackage{enumitem}           
\usepackage{graphicx}           
\usepackage{subcaption}         
\usepackage{float}              
\usepackage{txfonts}            
\usepackage{orcidlink}          
\usepackage{lipsum}             
\usepackage{lscape}             
\usepackage{placeins}           

\usepackage{hyperref}

\newcommand\setrow[1]{\gdef\rowmac{#1}#1\ignorespaces}

\begin{document}

\title{Blueberry and Green Pea galaxies live in low density environments}


\author{Maitrayee Gupta\inst{1}\thanks{maitrayee.gupta@asu.cas.cz}\orcidlink{0000-0003-0976-8932},
Ji\v{r}\'{i}~Svoboda\inst{1}\orcidlink{0000-0003-2931-0742},
Konstantinos~Kouroumpatzakis\inst{1}\orcidlink{0000-0002-1444-2016},
Nicolas Peschken\inst{2}\orcidlink{0000-0002-9925-2974},
Peter G. Boorman\inst{3}\orcidlink{0000-0001-9379-4716}
\and
Abhijeet Borkar\inst{1}\orcidlink{0000-0002-9807-4520}
        }


   \institute{Astronomical Institute of the Czech Academy of Sciences, Bo\v{c}n\'i II 1401, CZ-14100 Prague, Czech Republic,
            \and Department of Physics and Astronomy, University of Florence, Via Sansone, 1 - 50019 Sesto Fiorentino (FI), Italy
            \and Max Planck Institute for Extraterrestrial Physics, Giessenbachstrasse, 85741 Garching, Germany\\
            }

   \date{Received April 28, 2026}

 
\abstract{
\textbf{Context:} Little is currently known about the large-scale environments of Green Pea (GP) and Blueberry (BB) galaxies, which are low-mass, compact systems with extreme specific star-formation rates (sSFR). Their environments are inherently linked to their formation mechanism, and they may serve as crucial local analogues for high-redshift, reionizing galaxies. 

\textbf{Aim:} This paper aims to investigate the clustering properties of GPs and BBs, leveraging large-scale survey data to quantify
their spatial distribution relative to the broader galaxy population. 

\textbf{Methods:} We here investigate a sample of these galaxies, consisting of 339 GPs $\rm (0.1 < z \le 0.33)$ and 56 BBs $\rm (0 < z \le 0.1)$, whose clustering properties we analyse relative to an extensive control sample derived from the SDSS MPA-JHU DR8 catalogue, binned by stellar mass and sSFR. We use the number of neighbours within a 5 Mpc radius as a proxy for environmental density, i.e. clustering, and employ a pair-matching and bootstrapping methodology to ensure statistical robustness. 

\textbf{Results:} We observe that galaxy clustering depends strongly on star-formation activity, with passive galaxies being more clustered than their high star-formation rate counterparts, with GPs and BBs lying at the extreme end of this relation, exhibiting the lowest neighbour counts among all subsamples. The nearest neighbours of BBs also tend to have lower masses than other classes of dwarf galaxies. 

\textbf{Conclusions:} GPs and BBs predominantly reside in isolated, low-density environments, suggesting that their intense starbursts are unlikely to be triggered by common environmental processes such as mergers or starburst cycles. Their low metallicities and weak clustering instead support scenarios in which recent starbursts are driven by internal processes or pristine gas accretion, reinforcing their role as nearby analogues of young, low-mass galaxies in the early Universe.
}


\titlerunning{Clustering of GPs and BBs}
\authorrunning{Gupta et al.}

\maketitle
   
\section{Introduction}

It has been well studied that the degree of galaxy clustering exhibits a strong dependence on galactic stellar mass. More massive galaxies are consistently found to be more strongly clustered than their lower-mass counterparts across a wide range of redshifts (e.g. \citealt{2006MNRAS.368...21L}; \citealt{2011ApJ...736...59Z}). This mass-dependent clustering is interpreted as a reflection of the hierarchical structure formation process, where more massive galaxies tend to form and reside within more massive dark matter halos \citep{2010gfe..book.....M}. Consequently, the clustering amplitude of galaxies as a function of their mass provides critical insights into the galaxy-halo connection, informing models of how galaxies populate dark matter halos and how their baryonic content assembles over cosmic time \citep{2009ApJ...696..620C}. Conversely, less massive and actively star-forming galaxies typically exhibit weaker clustering, indicating a preference for lower-density regions or satellite positions within dark matter halos (\citealt{2008MNRAS.385..147D}; \citealt{2010ApJ...710L...1V}). These observed trends reflect the complex interplay between galaxy assembly, environmental influences (such as mergers and gas stripping), and the feedback mechanisms that regulate star formation over cosmic time.

Furthermore, the clustering of galaxies plays an equally pivotal role in modulating their star-formation rate (SFR), offering a distinct observational window into the physical mechanisms that shape galaxy evolution. Numerous studies have demonstrated that, at a fixed stellar mass, galaxies with higher SFRs (typically blue, star-forming galaxies) tend to be less clustered than galaxies with lower SFRs (typically red, quiescent galaxies; e.g. \citealt{2004MNRAS.353..713K, 2008ApJ...672..153C}; \citealt{2013ApJ...767...89M, 2020MNRAS.493.5987O}). The dependence of the SFR on the local environment density has been further investigated and confirmed in cosmological simulations \cite[e.g.][]{2014ApJ...788..133T}, including IllustrisTNG \citep{2019MNRAS.489..339H}, SIMBA \citep{2024MNRAS.528.4393G} and Magneticum \citep{2024A&A...683A..57R}. This differentiation suggests that star-forming galaxies preferentially inhabit lower-density environments or the outskirts of dark matter halos, while quiescent galaxies are more prevalent in dense regions where processes like mergers, gas stripping, or strangulation can quench star formation \citep{2014ApJ...782...33L}. Therefore, SFR-dependent clustering studies are crucial for understanding the environmental processes that trigger and terminate star formation, and for tracing the co-evolution of galaxies with their host dark matter halos (e.g. \citealt{2010ApJ...721..193P}; \citealt{2018MNRAS.475.3730C}).

Within the diverse tapestry of galaxy populations, Green Pea (GP) and Blueberry (BB) galaxies represent particularly intriguing classes. First identified through the Galaxy Zoo project, GPs are characterised by their compact size (typical radius of 0.33 kpc; \citealt{2021ApJ...914....2K}), low stellar masses $\rm (M_{\star}\sim 10^{8.5}-10^{10}M_{\odot})$, extremely high specific star-formation rates (sSFR; $\rm sSFR \equiv SFR/M_{\star}$), and strong $\rm [O III]$ emission lines, giving them their distinctive green hue in composite SDSS images in the redshift range $\rm 0.112 < z < 0.360$ \citep{2009MNRAS.399.1191C}.

BB galaxies, which are identified by their compact sizes and blue optical colours, tend to occupy even lower metallicity regimes and display even higher sSFR, and are considered their lower-redshift (z < 0.1), lower-mass, and often bluer counterparts \citep{2017ApJ...847...38Y, 2024A&A...688A.159K}. Both populations are considered vital local analogues for understanding the high-redshift, Lyman-alpha emitting galaxies thought to be responsible for cosmic reionization (\citealt{2022A&A...665L...4S}; \citealt{2023ApJ...942L..14R} and review by \citealt{2025ARA&A..63...45J}). Their intense, concentrated starburst activity in relatively low-mass systems raises questions about their formation triggers and their typical large-scale environments. 
GPs and BBs likewise present an opportunity to gain a unique perspective on the roles of Active Galactic Nuclei (AGN) versus star-formation in driving cosmic reionization by performing detailed X-ray and radio studies on nearby analogues of high-redshift galaxies \citep{2019ApJ...880..144S, 2024A&A...691A..27A, 2024A&A...687A.137B}. Recent developments include identifying an AGN in a GP galaxy \citep{2025ApJ...988..157B}. Studying GPs and BBs could also shed light on the recently-discovered "little red dots" (LRDs), which are a new class of high-redshift ($\rm z > 4$) compact galaxies discovered by the James Webb Space Telescope that are red in the rest-frame optical and blue in the rest-frame UV with V-shaped spectral energy distributions \citep[SEDs;][]{2023ApJ...956...61A, 2023ApJ...942L..14R, 2024ApJ...963..128B, 2025ApJ...978...92L, 2025ApJ...986..126K}. These LRDs are understood to be the high redshift counterparts of GPs and BBs \citep{2025ApJ...980L..34L}.

This paper aims to investigate the clustering properties of GPs and BBs, leveraging large-scale survey data to quantify their spatial distribution relative to the broader galaxy population. We herein study the environmental density of these populations using the number of neighbours within a radius of 5 Mpc as a proxy for their clustering. We will explore how their clustering varies, if at all, compared to other galaxies matched in stellar mass but with lower SFR. Understanding the environments in which GPs and BBs thrive is crucial for constraining their formation mechanisms – whether they are driven by mergers, specific gas accretion modes, or other processes – and for solidifying their role as proxies for understanding galaxy formation and feedback in the early Universe.

The work presented here is organised as follows: in Section \ref{sect:sample} we describe the procedure we used to select our sample; in Section \ref{sect:Analysis} we present the methodology of the clustering study; in Section \ref{sect:discussion} we discuss our results, and the main findings are summarised in Section \ref{sect:conclude}.
Throughout the paper, we assume a $\Lambda$CDM cosmology with 
$\rm H_0 = 70 \ {\rm km} \ {\rm s}^{-1} \ {\rm Mpc}^{-1}$, $\rm \Omega_m=0.3$, and $\rm \Omega_{\Lambda}=0.70$. \\ \\

\begin{figure}
\centering
\includegraphics[width=0.99\linewidth]{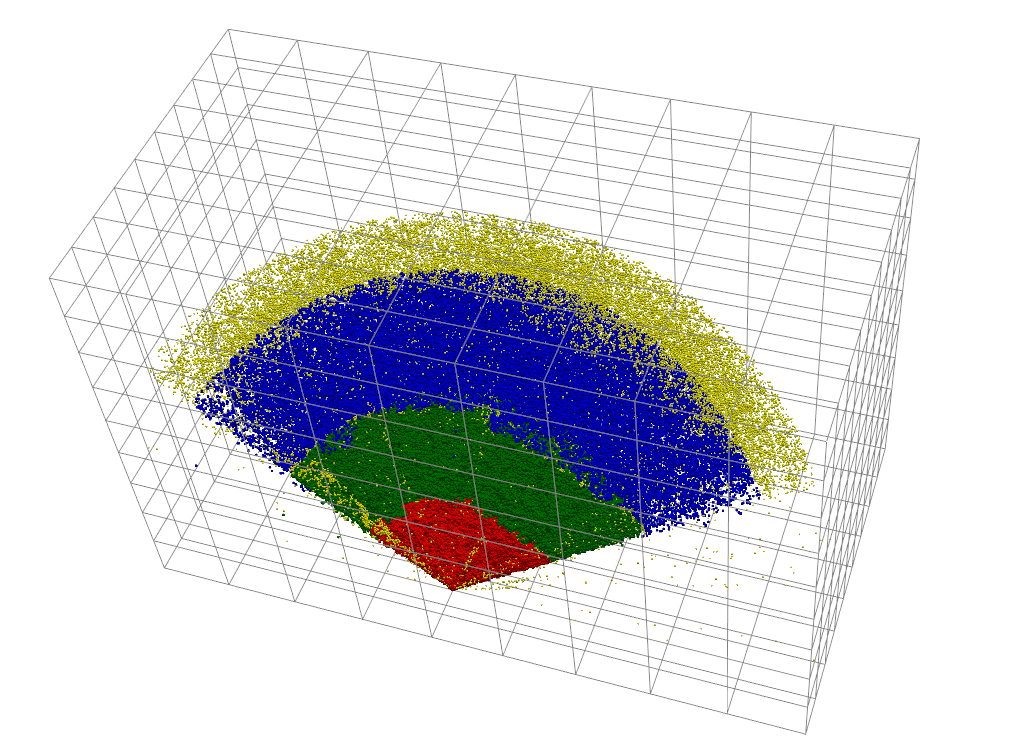}
\caption{3D distribution showing the parent sample of galaxies. The selected objects in the three studied redshift ranges are shown in red ($\rm 0 < z < 0.1$), green ($0.1 < z < 0.2$), and blue ($0.2 < z < 0.33$) Objects outside of these redshift ranges are shown in yellow.}
\label{fig:clustering_parent_full}
\end{figure}

\begin{figure}
\centering
\captionof{table}{Definition of the 9 galaxy subsets based on their galaxy masses and sSFR, showing the number of objects in each class. We binned each sample by galaxy mass bins and sSFR with three bins for each, resulting in 9 subsets. These subsets were then labelled as (1) Passive Dwarf, (2) Normal Dwarf, (3) Starburst Dwarf, (4) Passive Average, (5) Normal Average, (6) Starburst Average, (7) Passive Massive, (8) Normal Massive, and (9) Starburst Massive. 
}
\includegraphics[width=\linewidth]{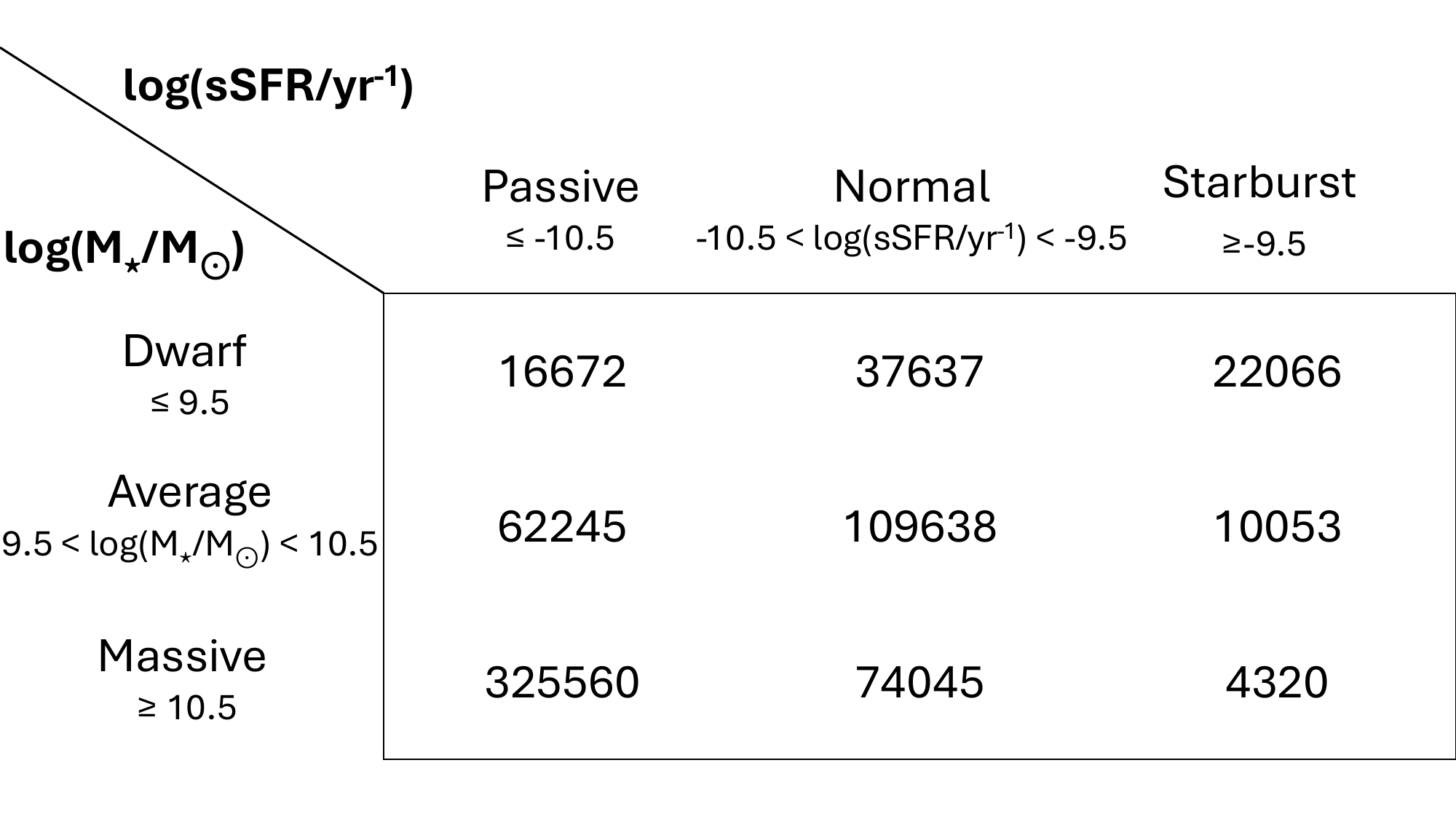}
\label{tbl_define}
\end{figure}

\section{Samples and Data Analysis}
\label{sect:sample}
\subsection{Parent Sample}

To establish a baseline for comparison, we created a parent sample from the SDSS galaxy sample MPA-JHU (Max Planck Institute for Astrophysics and the Johns Hopkins University) DR8 catalogue \citep{2003MNRAS.341...33K, 2004MNRAS.351.1151B}, which has galaxy spectral properties for all of the galaxies in the SDSS DR8 catalogue \citep{2011ApJS..193...29A}, including stellar mass ($\rm M_\star$) and sSFR. The sSFR was estimated by combining emission line measurements from within the fibre where possible, and aperture corrections are done by fitting models by \cite{2005MNRAS.362...41G} and \cite{2007ApJS..173..267S}. By retaining confirmed galaxies from that sample (based on the classification of the spectrum \cite{2012AJ....144..144B}), we created a parent sample of 1068446 objects. 
\begin{figure*}
\centering
\includegraphics[width=0.99\linewidth]{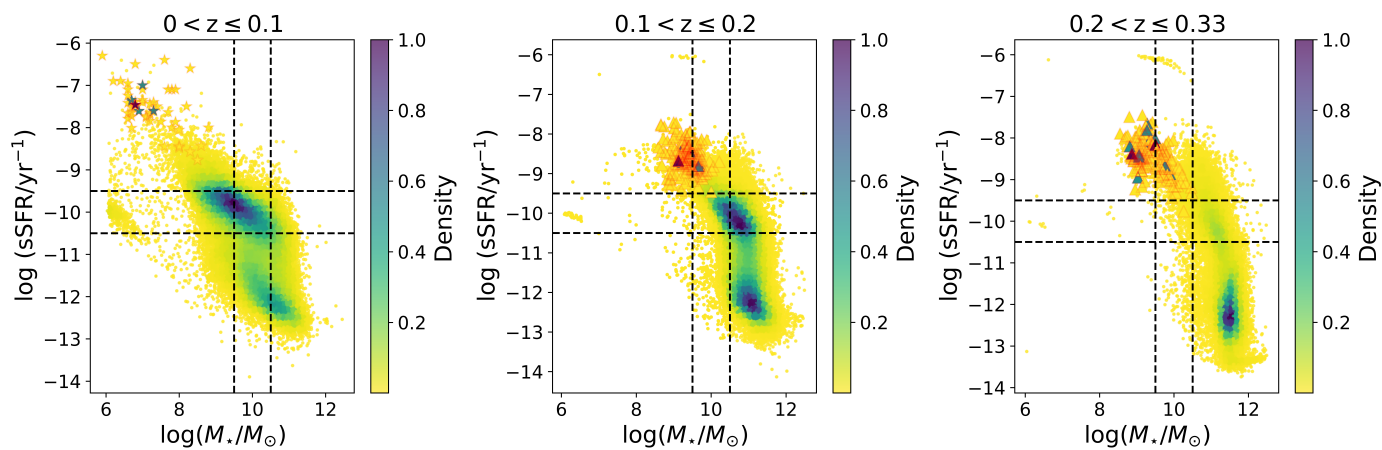}
\caption{Scatter plot showing the number density distribution of the stellar masses vs. sSFR for the 3 redshift samples. BBs are represented with stars and GPs are represented with triangles. The horizontal dashed lines delineate the 3 sSFR bins and the vertical dashed lines the 3 stellar mass bins.}
\label{fig:scatter_plot_redshift}
\end{figure*}





We applied a redshift cutoff of $\rm 0 < z \le 0.33$ to this sample, as our BB sample is in the redshift range $\rm 0 < z \le 0.1 $ and the GP sample is in the redshift range $\rm 0.1 < z \le 0.33$. This was done to avoid biases introduced due to cosmological evolution effects, and is shown in Figure \ref{fig:clustering_parent_full}.
 
Objects without valid measurements of galaxy mass or sSFR estimates in the MPA-JHU sample were removed from further analysis. Additionally, to reduce the error and select reliable estimates, we eliminated objects with a high sSFR error, (e.g. \citealt{2014ApJS..214...15S}, \citealt{2023A&A...673A..16K}), which we defined as those with one sigma error greater than $\rm 10\%$.

We divided the remaining galaxies into 3 sSFR bins:
\begin{itemize}
  \item Passive: $\rm -10.5 \le log(sSFR/yr^{-1})$
  \item Normal: $\rm -10.5 < log(sSFR/yr^{-1}) < -9.5$
  \item Starburst: $\rm log(sSFR/yr^{-1}) \ge -9.5$
\end{itemize}

We similarly divided these galaxies into 3 galaxy mass bins:
\begin{itemize}
  \item Massive: $\rm log(M_{\star}/M_{\odot}) \ge 10.5$
  \item Average: $\rm 9.5 < log(M_{\star}/M_{\odot}) < 10.5$
  \item Dwarf: $\rm log(M_{\star}/M_{\odot})\le 9.5$
\end{itemize}

Applying both categorisations divided our sample into 9 subsamples, seen in Table \ref{tbl_define}. These were labelled as (1) Passive Dwarf, (2) Normal Dwarf, (3) Starburst Dwarf, (4) Passive Average , (5) Normal Average, (6) Starburst Average, (7) Passive Massive, (8) Normal Massive, and (9) Starburst Massive, as illustrated in Table \ref{tbl_define}. These subsamples were used as our control sample in comparisons against GPs and BBs. Table \ref{tbl_obj} lists the number of objects in each subsample by redshift.


\begin{figure*}
\centering
\includegraphics[width=0.75\linewidth]{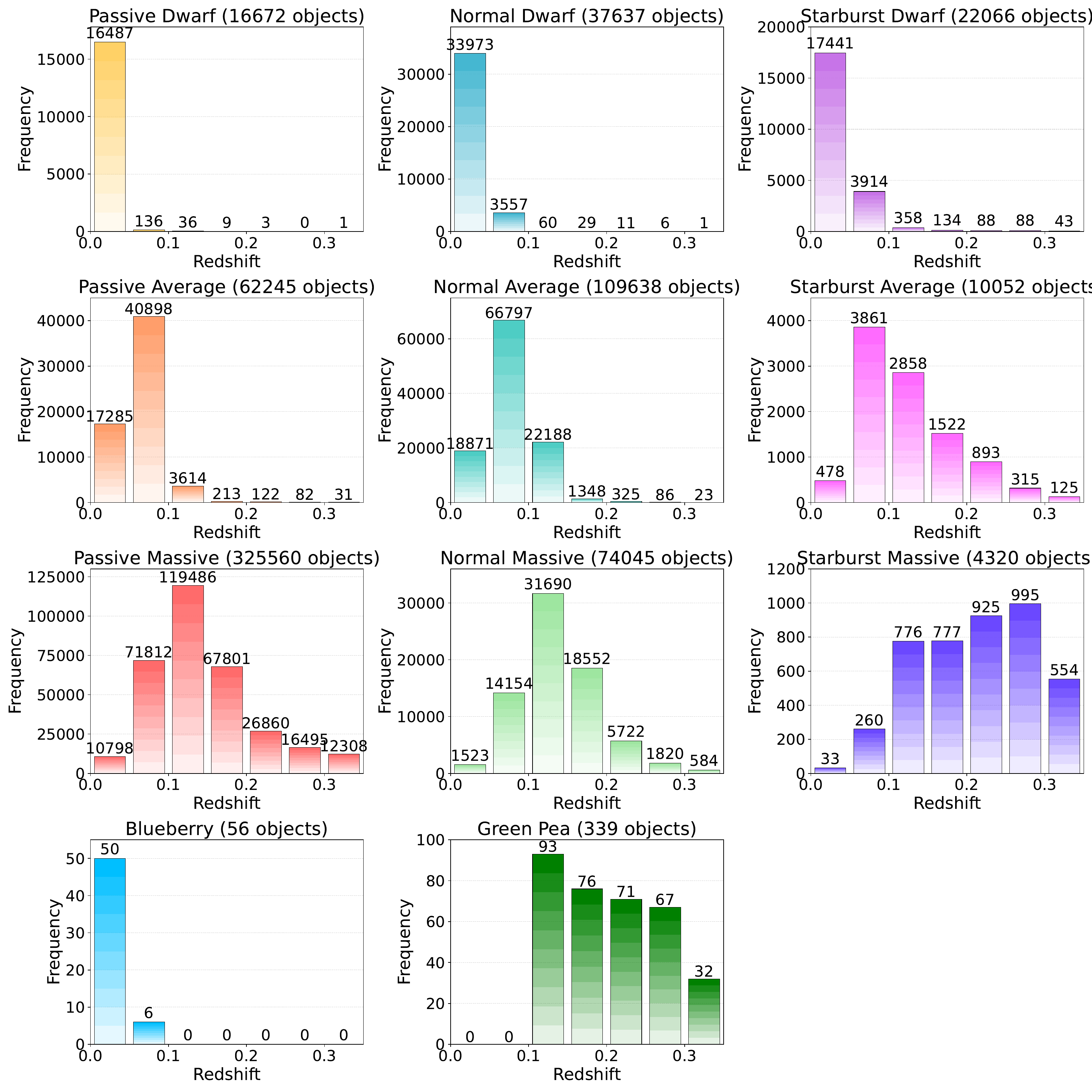}
\caption{Histogram showing the redshift distribution of the 9 control samples and the GP and BB samples. The number of objects in each bin is shown above each bar in the histogram, with the total number of objects overall above each plot.}
\label{fig:histogram_plot_redshift}
\end{figure*}

\subsection{Green Pea galaxy sample}

We constructed our sample of GPs from a number of sources. Firstly, we included all objects in the \cite{2009MNRAS.399.1191C} sample, which consists of 80 GPs selected from the SDSS Data Release 7 (DR7)
spectroscopic sample \citep{2009ApJS..182..543A}. Our next source was the GPs identified in \cite{2022AJ....163..150K}, which were taken from the Gems of the Galaxy Zoos project \citep{2008MNRAS.389.1179L}. We selected a total of 19 galaxies from this source, as there was some overlap with the sample identified in our first source.

To further supplement this, we selected 240 GP-like objects from our parent sample based on the criteria in \cite{2011ApJ...728..161I}. These include:\\

\begin{enumerate}[nosep]
\item The extinction-corrected luminosity of the $\rm H \beta$ emission line is greater than $\rm 3 \times 10^{40} erg\,s^{-1}$. 
\item The equivalent width of the $\rm H \beta$ emission line is $\rm \ge 50 \AA$. 
\item Only galaxies with a well-detected $\rm [O_{III}] \lambda 4363$ emission
line in their spectra, i.e. those with a flux error less than 50\% of the line flux.
\item Galaxies with obvious evidence of Seyfert 2 spectral features are excluded (such features include strong $\rm [Ne_{V}] \lambda 3426$, $\rm [He_{II}] \lambda 4686$, $\rm [O_{I}] \lambda 6300$, $\rm [S_{II}] \lambda 6717$, and $\rm \lambda 6731$ emission lines). Previous studies have shown that BPT diagrams, along with other standard techniques, do not work reliably as a means to identify AGN in these star-forming dwarfs (e.g. \citealt{2006MNRAS.371.1559G, 2019ApJ...870L...2C, 2020MNRAS.492.2268B, 2022ApJ...931...44P, 2024MNRAS.528.5252M, 2024ApJ...971...68W} etc.). Moreover, as the primary goal of this work is to study the environmental properties of BBs and GPs as populations, the presence (or lack thereof) of AGN in these galaxies does not have an impact on the main analysis. While there may be some AGN contamination in our sample, previous studies on identifying AGN in BB and GP sources (e.g. \citealt{ 2009MNRAS.399.1191C, 2019ApJ...880..144S, 2025ApJ...988..157B}) suggest that only a very small fraction of such sources have AGN, and therefore our results would not be strongly affected.
\item Visual confirmation that the selected objects are compact at low redshifts and unresolved at high redshifts (see \cite{2011ApJ...728..161I} for more details on this selection criterion). This process eliminates "Extended Peas" from the recently-discovered population by \cite{2025arXiv251108678S}, which may contribute to a lack of merger signatures.

\item For our analysis, we additionally restricted our selections to only those in the redshift range 0.1 to 0.33. 
\end{enumerate}

\begin{figure}
    \centering
\includegraphics[width=0.9\linewidth]{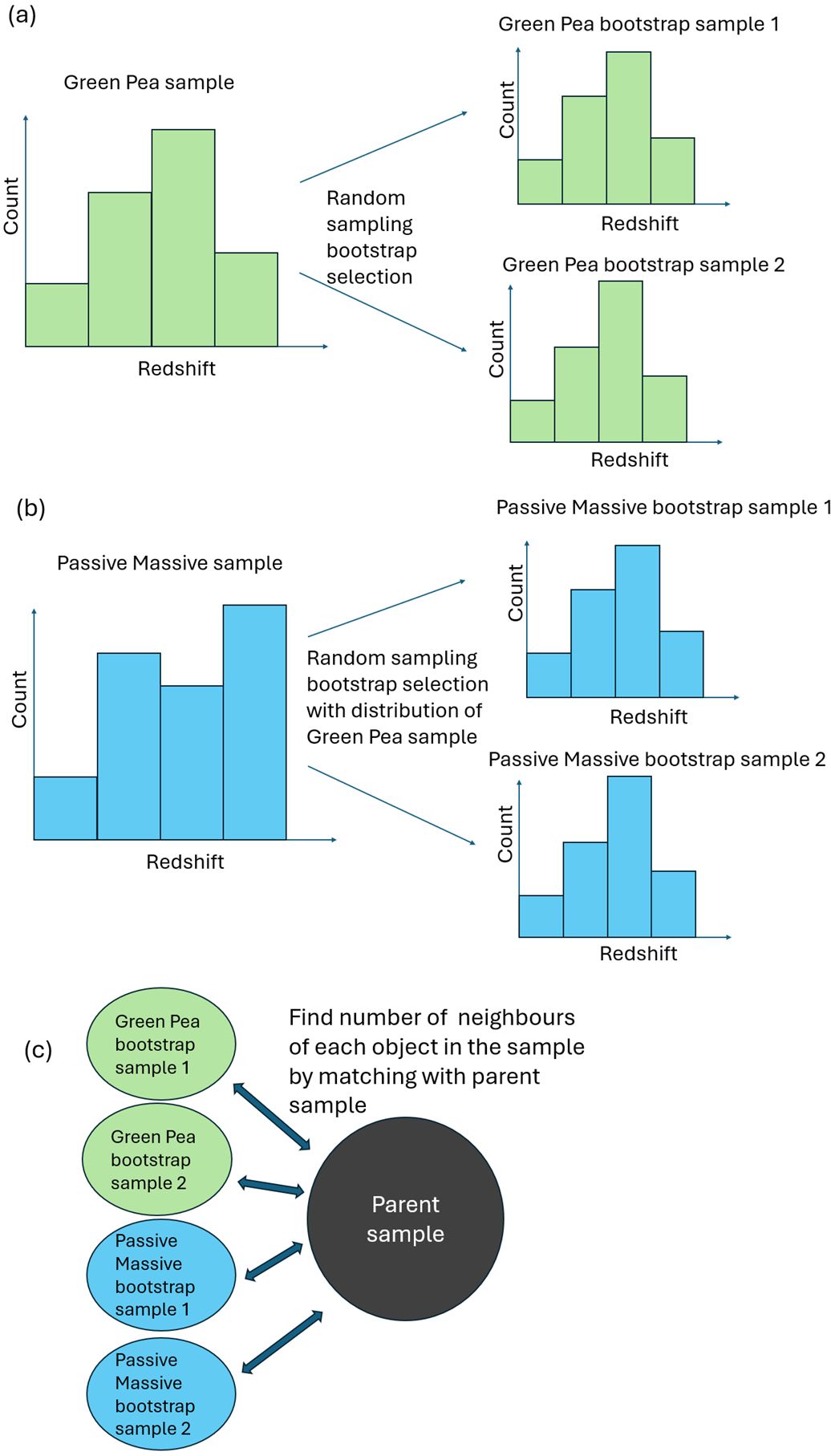}
    \caption{Pair-matching and bootstrapping mechanism. (a) The GP sample (i.e., the smaller among the samples being compared), is used as the template dictating the desired shape of the histogram for the subsamples. Multiple such subsamples are produced by random selection from the GPs to produce 500 bootstrapped subsamples. (b) For each other class of objects, we generate subsamples of that class with a similar redshift distribution to the GP sample. We do this by randomly selecting objects from that class until we have a subsample whose redshift distribution matches that of the GP sample. Using this method, we produce 500 bootstrapped subsamples for each of the object classes. (c) Each of the constructed subsamples is matched against the parent sample; all galaxies within 5 Mpc are noted down as neighbours and considered for further study.}
    \label{fig:pair_matching}
\end{figure}

Our final GP sample contains 339 objects, with 169 belonging to the redshift range $0.1 < z \le 0.2$ (GP Sample 1) and 170 to the range $0.2 < z \le 0.33$ (GP Sample 2).

\subsection{Blueberry galaxy sample}

Our sample of 56 BB galaxies was selected from \citet{2017ApJ...847...38Y}, who identified BBs using photometric data from SDSS Data Release 12 (DR12; \citealt{2015ApJS..219...12A}), and from \citet{2024A&A...688A.159K}, who also used photometric data from Pan-STARRS \citep{2016arXiv161205560C}.
We restricted our analysis to only those objects in the redshift range less than 0.1. 

\begin{table}
\caption{Number of objects in each redshift range.}
\label{tbl_obj}
\begin{center}
\resizebox{\columnwidth}{!}{
 \begin{tabular}{c c c c} 
 \hline
 Object Class &  $0 < z \le 0.1$ & $0.1 < z \le 0.2$ & $0.2 < z \le 0.33$ \\ 
 \hline\hline
Passive Massive & 82610 & 187287 & 55663 \\
Passive Average & 58183 & 3827 & 235 \\
Passive Dwarf & 16623 & 45 & 4 \\
Normal Massive & 15677 & 50242 & 8126 \\
Normal Average & 85668 & 23536 & 434 \\
Normal Dwarf & 37530 & 89 & 18 \\
Starburst Massive & 293 & 1553 & 2474 \\
Starburst Average & 4339 & 4380 & 1334 \\
Starburst Dwarf & 21355 & 492 & 219 \\
BBs & 56 & - & - \\
GP Sample 1 & - & 169 & - \\
GP Sample 2 & - & - & 170 \\
 \hline
\end{tabular}
}
\end{center}
\end{table}

\begin{table}
\caption{
Results of K-S test comparing redshift distributions for the redshift range $0 < z \le 0.1$: we compare the original BB sample with the original galaxy samples as well as a pair-matched sample. $\rm p \ge 0.05$ indicates that we cannot reject the hypothesis that the two redshift samples share the same distribution and $\rm p < 0.05$ indicates that they are different.}
\label{ks_test_pair_match}
\begin{center}
\resizebox{\columnwidth}{!}{
 \begin{tabular}{c c c} 
 \hline
Sample compared  & P-value of KS test  & P-value of KS test \\
with the BB & comparing with&comparing with \\
sample& Original sample& pair-matched sample\\
 \hline\hline
Passive Massive & 1.6e-42 & 0.92 \\
Passive Average & 8.4e-23 & 0.90  \\
Passive Dwarf & 7.7e-44 &  0.21 \\
Normal Massive & 3.5e-48 &  0.78\\
Normal Average & 1.2e-29 & 0.93  \\
Normal Dwarf & 1.8e-09 & 0.81 \\
Starburst Massive & 2.1e-39 & 0.63 \\
Starburst Average & 1.1e-44 & 0.84 \\
Starburst Dwarf & 1.9e-13 & 0.89  \\
\hline
\end{tabular}
}
\end{center}
\end{table}


\begin{figure*}
\centering
\includegraphics[width=0.99\linewidth]{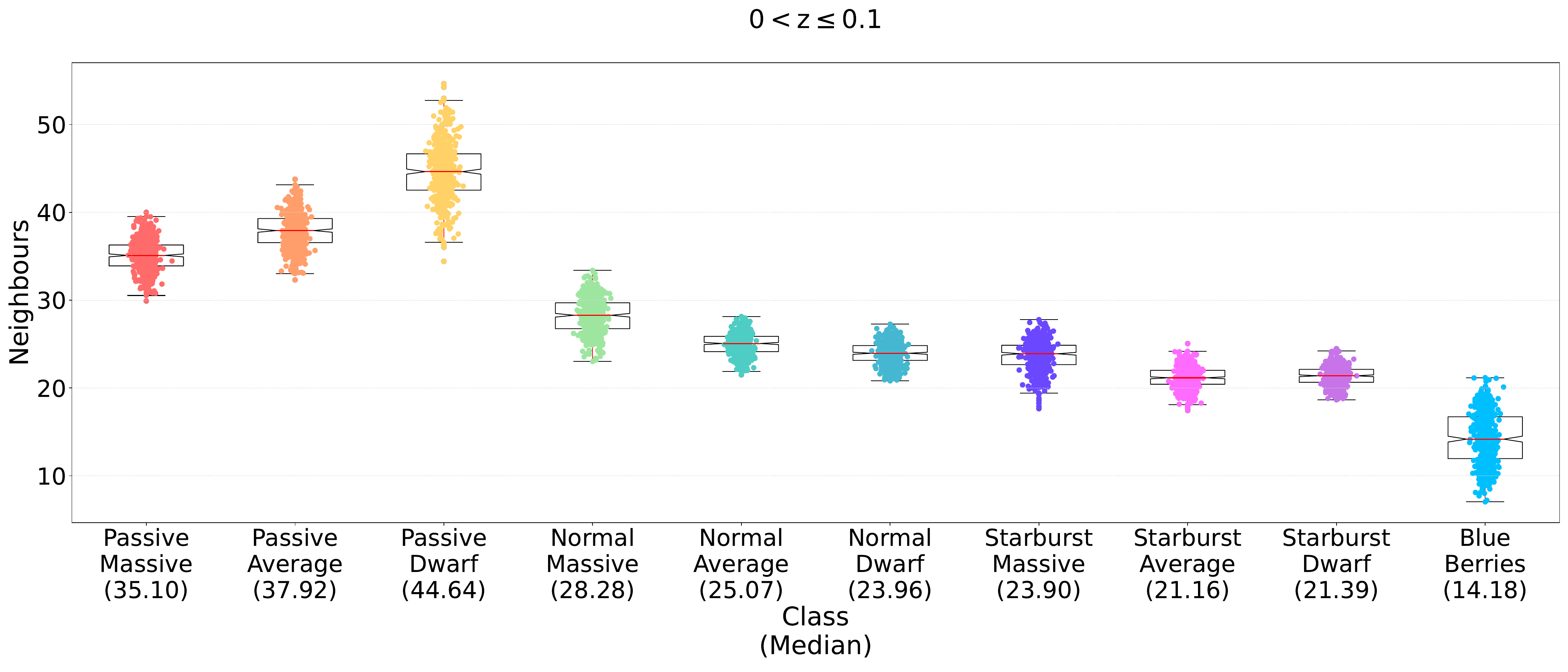}
\includegraphics[width=0.99\linewidth]{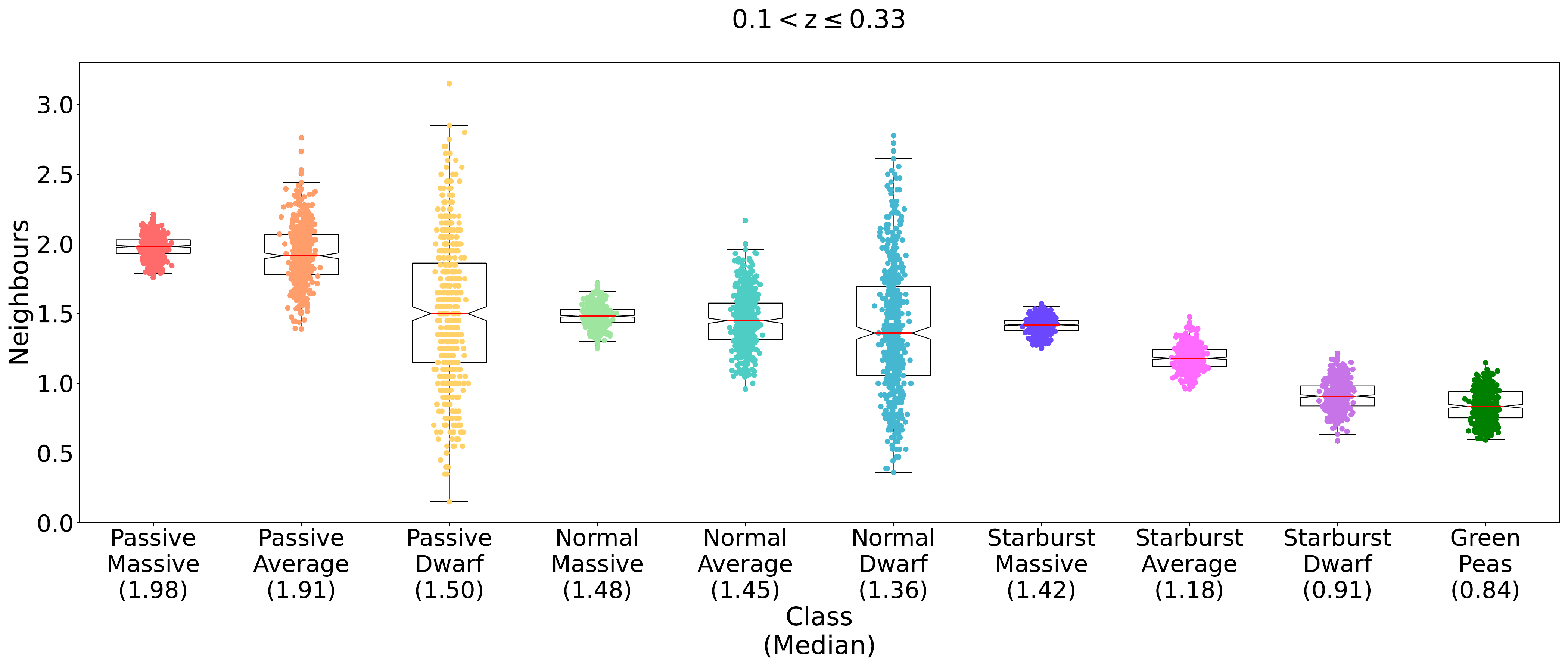}
\caption{Box plot showing the average number of neighbours within 5 Mpc for the 500 bootstrapped samples for each class of object in the redshift range $0 < z \le 0.1$ (top panel) and redshift range $0.1 < z \le 0.33$ (bottom panel). Each point is the mean of the number of neighbours for a bootstrapped sample, resulting in 500 points per class. The median number of neighbours for each class is shown with a red line, and is recorded numerically below the class names.}
\label{fig:clustering_parent_full_range14}
\end{figure*}

\subsection{Analysis}
\label{pair_match}

Table \ref{tbl_obj} lists the total number of objects in each subsample, including the BB Sample, GP Sample 1, and GP Sample 2, by redshift. Figure \ref{fig:scatter_plot_redshift} shows the distribution of objects in the three redshift ranges. Figure \ref{fig:histogram_plot_redshift} shows the redshift distribution of the 9 control samples, the BB samples, and the GP samples. As seen in the figure, the subsamples occupy a wide range of redshift distributions. 

Since the GP and BB samples are much smaller than the parent sample and have very different redshift distributions, a straightforward analysis without adjustments would immediately present redshift effects that would bias the results. Therefore, to avoid this bias due to selection effects, we match the redshift distributions of all control subsamples to those of BBs and GPs for the corresponding redshift ranges, respectively. To accomplish this, we used a pair-matching technique wherein we created a redshift histogram of the GP or BB sample and then selected from the nine non-GP, non-BB subsamples a proportional number of objects such that their histogram resembled that of the GP or BB sample. This process is illustrated in Figure \ref{fig:pair_matching}.

Table \ref{ks_test_pair_match} shows the results of the K-S test comparing the redshift distributions of the objects in one of the redshift ranges; $0 < z < 0.1$. We compare the original BB sample with the original galaxy samples as well as a pair-matched sample. A value of $\rm p \ge 0.05$ indicates that we cannot reject the hypothesis that the two redshift samples share the same distribution, while $\rm p < 0.05$ indicates that they are different. Results for other redshift ranges are similar and omitted for brevity. These tests suggest that the pair-matching approach successfully minimises selection biases while preserving the statistical integrity of the clustering analysis.

Additionally, in order to improve the statistical significance of the clustering results, we made use of bootstrapping, i.e. the above pair-matching process is repeated to generate 500 bootstrapped samples for each class of objects; this bootstrapping process was done with replacement. Figure \ref{fig:pair_matching} demonstrates the process. The figure illustrates this only for two samples for brevity. In each of the three redshift ranges, the ten galaxy samples (i.e. 9 control samples and GP or BB sample as appropriate) were matched with the parent sample, and all galaxies within 5 Mpc were considered neighbours and taken for further study. In the following section we will present the clustering properties, distances, and masses of these neighbours. 

To quantify the clustering of each galaxy class, we adopted the number of neighbours within a fixed radius of 5 Mpc as an environmental-density proxy. This simple statistic provides a stable measure of local overdensity even for relatively small samples such as the GP and BB populations, for which a traditional two-point correlation function (2PCF) would be dominated by Poisson noise. The 5 Mpc aperture was chosen as a compromise scale, large enough to probe regions beyond the typical virial radius of galaxy groups and clusters $({\sim} 1$-$2$~Mpc) while staying small enough to remain sensitive to variations in the local large-scale structure. This scale has been widely used in environmental studies to separate field from group or cluster galaxies (e.g. \citealt{2004MNRAS.353..713K})

\begin{figure*}
\centering
\begin{subfigure}[b]{0.32\textwidth}
\centering
\includegraphics[width=\textwidth]{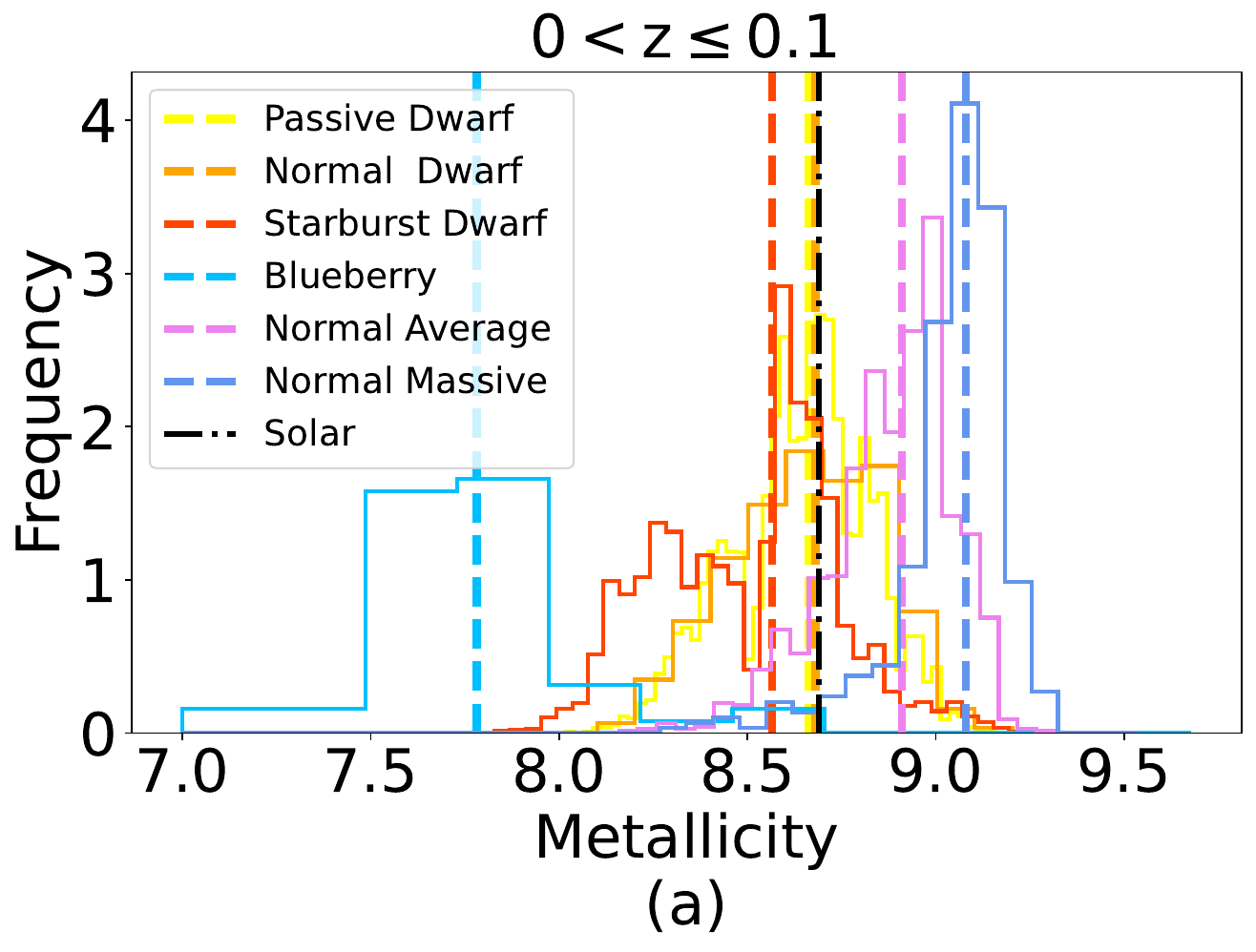}
\label{fig:hist_metallicty_range1}
\end{subfigure}%
\begin{subfigure}[b]{0.32\textwidth}
\centering
\includegraphics[width=\textwidth]{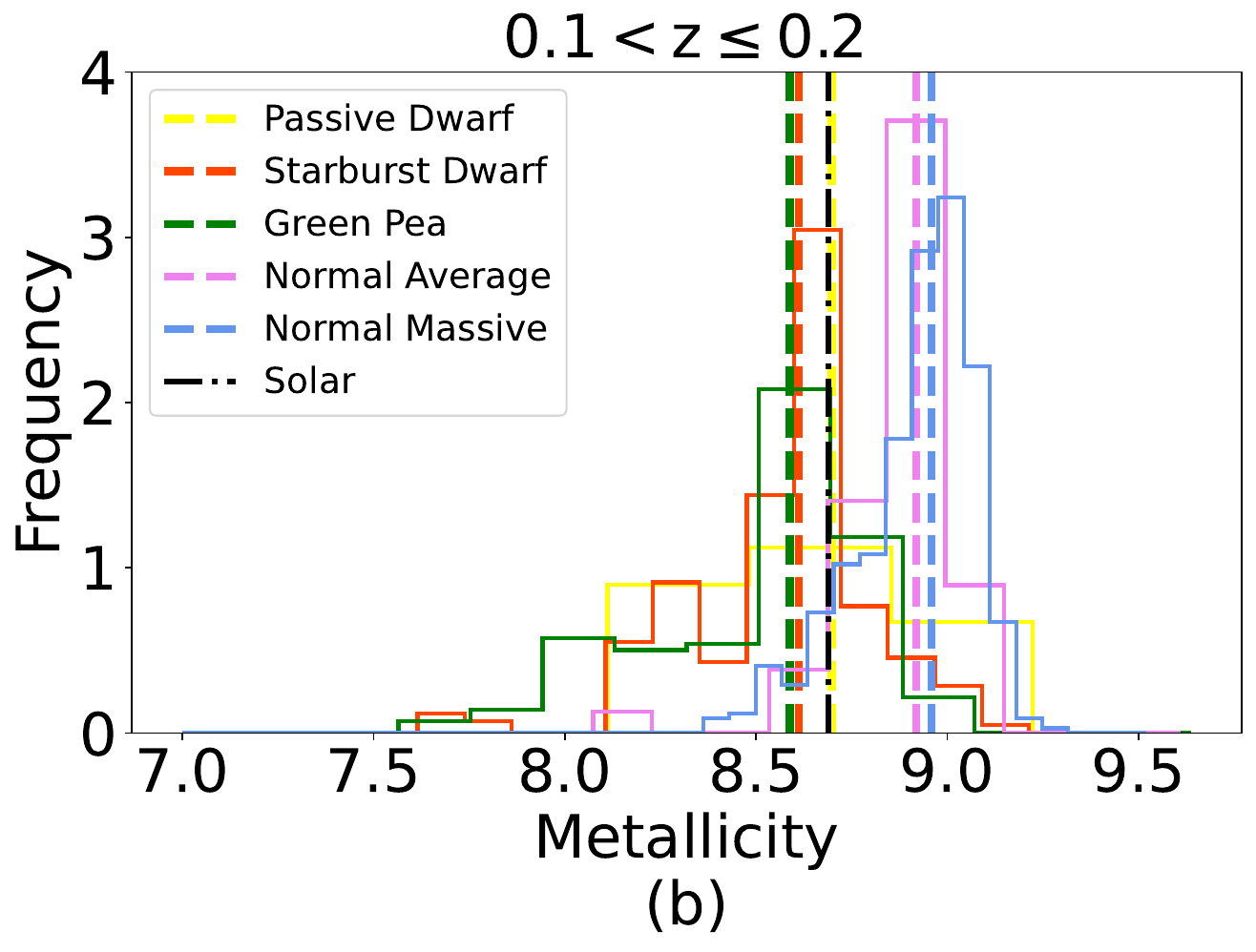}
\label{fig:hist_metallicty_range2}
\end{subfigure}%
\begin{subfigure}[b]{0.32\textwidth}
\centering
\includegraphics[width=\textwidth]{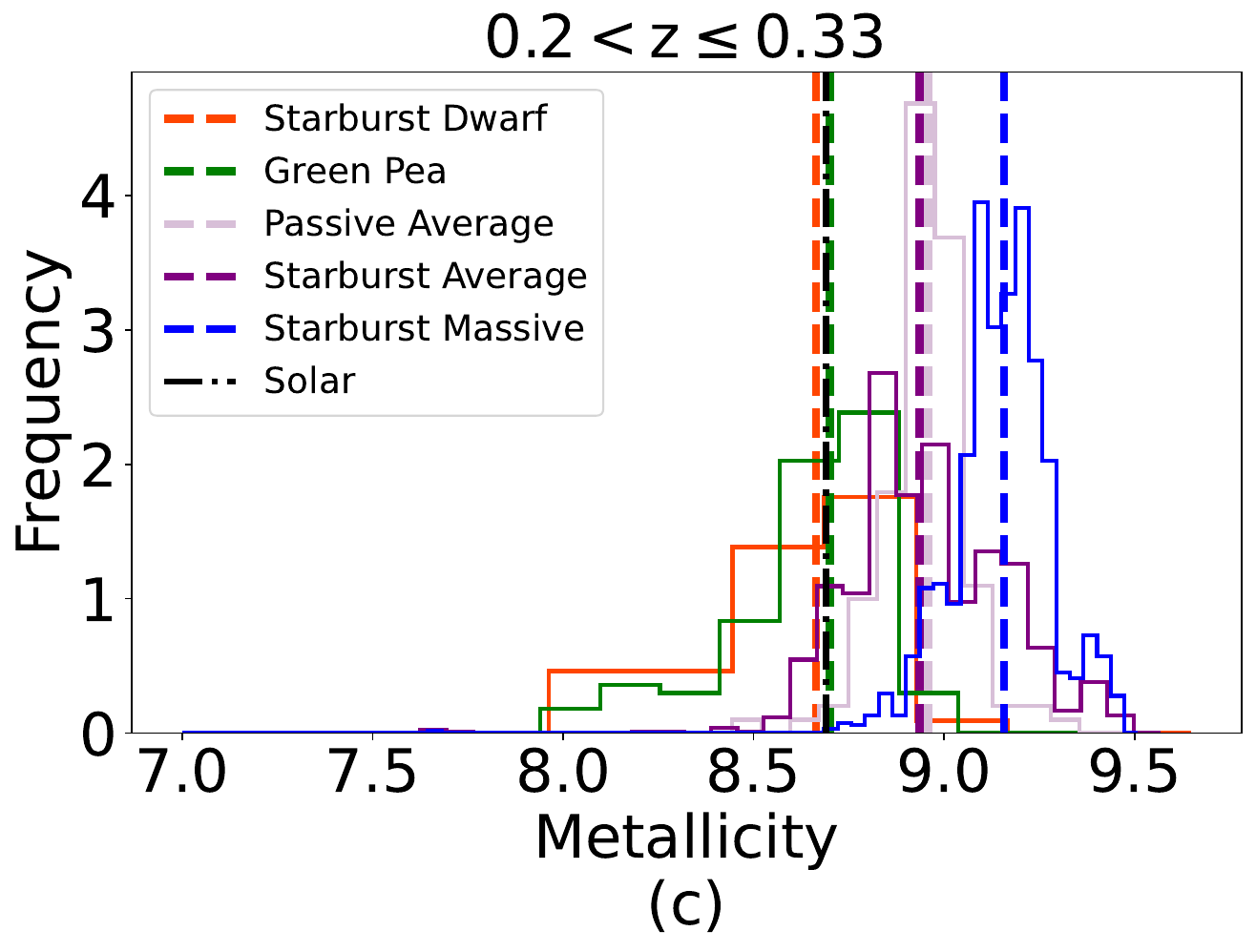}
\label{fig:hist_metallicty_range3}
\end{subfigure}
\vspace{-15pt} 
\caption{Distribution of the metallicity of objects in different redshift ranges, (a) $0 < z \le 0.1$, (b) $0.1 < z \le 0.2$ and (c) $0.2 < z \le 0.33$. Not all classes of objects are shown for readability. The median values shown in the plots are listed in Table \ref{tbl_metal}. The black line shows the solar metallicity, $\rm 12+log(O/H) = 8.69$ \cite{2009ARA&A..47..481A}}
\label{fig:hist_metallicty_ranges}
\end{figure*}

\begin{table*}
\caption{Post-Hoc Dunn Analysis to identify which specific pairs of classes have significantly different clustering distributions. The table shows the p-value resulting from a test of the null hypothesis that the distributions of the two groups being compared are equal. A p-value less than or equal than 0.05 rejects the null hypothesis; this shows that our results are statistically significant.}
\label{tbl_post_hoc}
\begin{center}
 \begin{tabular}{c c c c c} 
 \hline
 Object Class &  $0 < z \le 0.1$ & $0.1 < z \le 0.2$ & $0.2 < z \le 0.33$  & $0.1 < z \le 0.33$\\ 
 & compared against &compared against &compared against & compared against\\
  & BB Sample & GP Sample 1 & GP Sample 2& Complete GP Sample \\
\hline\hline
\setrow{\bfseries}Passive Massive & 0 & 0 & 1.85e-187 & 0 \\
\setrow{\bfseries}Passive Average & 0  & 0 & 2.46e-119 & 0 \\		
\setrow{\bfseries}Passive Dwarf & 0 & 6.84e-223 & 6.55e-62 &	2.55E-132 \\
\setrow{\bfseries}Normal Massive & 1.86e-221 & 2.14e-202 & 1.06e-75 & 6.12E-154\\
\setrow{\bfseries}Normal Average & 4.30e-132  & 1.44e-103 & 9.37e-21 & 5.07E-127 \\
\setrow{\bfseries}Normal Dwarf & 6.19e-90 & 1.51e-89 & 6.55e-62 & 6.43E-95 \\
\setrow{\bfseries}Starburst Massive & 1.05e-81  & 1.09e-142 & 8.92e-28 & 6.13E-106 \\
\setrow{\bfseries}Starburst Average & 3.93e-17  & 6.86e-26 & 0 & 3.26E-30 \\
\setrow{\bfseries}Starburst Dwarf & 2.34e-19  & 0.01 & 6.94e-16	& 0.04 \\
\hline
\end{tabular}
\end{center}
\end{table*}

\begin{figure}
\centering
\includegraphics[width=0.99\linewidth]{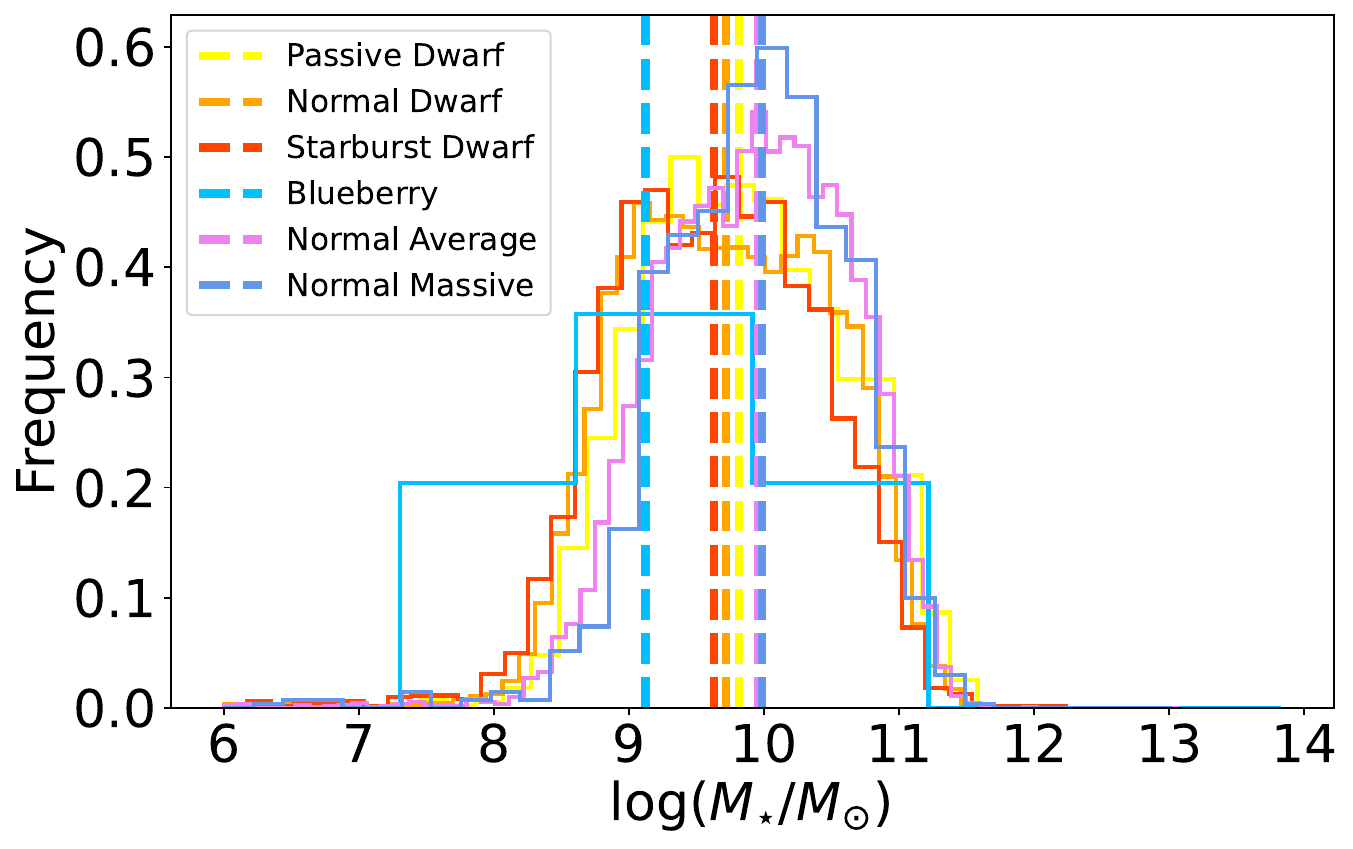}
\caption{Histogram showing mass of the nearest neighbour for each class of objects in the redshift range $0 < z \le 0.1$ up to a distance of 1 Mpc, with some classes of objects omitted for readability. The median values shown in the plot are listed in Table \ref{tbl_mass}.}
\label{fig:hist_nearest}
\end{figure}

\section{Results}
\label{sect:Analysis}

We present here our observations for each sample in our study regarding their clustering properties, as well as the mass of and distance to their nearest neighbour and their metallicity. 

\subsection{Clustering Properties}
\label{sect:cluster}

As stated above, we adopt the number of neighbours within 5 Mpc of each selected galaxy as a proxy for the clustering properties of our samples, and use this number as a metric to understand the density of each galaxy's environment. 

The mean number of neighbours for each of the 500 bootstrapped samples is presented as a box plot in Figure \ref{fig:clustering_parent_full_range14}, with the top panel showing the results for redshift range $0 < z \le 0.1$, which contains the BBs, and the bottom panel showing the results for range $0.1 < z \le 0.33$, which contains the GPs. For clarity, the combined $0.1 < z \le 0.33$ range is shown in Figure \ref{fig:clustering_parent_full_range14}, while the individual redshift bins  $0.1 < z \le 0.2$ and $0.2 < z \le 0.33$ are presented in the appendix (Figure \ref{fig:clustering_parent_full_range23}).

In each of those figures, each coloured point represents the mean number of neighbours for a bootstrapped sample. The median of these 500 means is indicated by a red line in the box, while the mean of those means is shown as a green triangle. In each plot, the extremities of the boxes represent the 25th and 75th percentiles of the distributions.

In both redshift ranges, these figures clearly show a continuous trend where the number of neighbours decreases with increasing sSFR. This is in agreement with previous observations that field galaxies tend to have higher star formation rates and that passive galaxies tend to primarily reside in galaxy clusters \citep{2012ApJ...757...85G, 2024ApJ...971..111R, 2025ApJ...978...67M, 2025MNRAS.536..905G}. This trend holds true for massive galaxies, average galaxies, and dwarf galaxies, with GPs and BBs having the lowest numbers of neighbours amongst all studied populations. As these are among the most actively star-forming classes of galaxies in the modern Universe, this observation follows expected trends. We do not observe any statistically significant change in the number of neighbours based on the masses of the galaxies.

It is important to note that the differences in the median number of neighbours between Figures \ref{fig:clustering_parent_full_range14} and \ref{fig:clustering_parent_full_range23} can be attributed to lower overall densities of sources at larger distances due to expected observational bias.

We also note the metallicity distribution for the galaxies in our sample. We adopted the metallicities provided in the MPA-JHU catalogue, which were estimated using \citealt{2004ApJ...613..898T} and \citealt{2004MNRAS.351.1151B}. Figure \ref{fig:hist_metallicty_ranges} presents these histograms, in which we define the metallicity distribution as 
log[O/H] + 12, where the Solar metallicity is typically around 8.69 \cite{2009ARA&A..47..481A}, which we show as a black line in each plot. Figures \ref{fig:hist_metallicty_ranges} a, b and c show the distributions for redshift ranges $0 < z \le 0.1$, $0.1 < z \le 0.2$, and $0.2 < z \le 0.33$, respectively. 
In the following comparisons, the bins used in each histogram are based on Scott's rule \citep{10.1093/biomet/66.3.605}.
The medians of these distributions are displayed as dashed lines and can be found in Table \ref{tbl_metal}. We observe that the GPs and BBs tend to have very low metallicities, which agrees with previous findings in \cite{2009MNRAS.399.1191C} and \cite{2024A&A...688A.159K}. 
In Figure \ref{fig:hist_metallicty_ranges}c the dwarf passive results have poor statistical significance due to a small sample size (4 galaxies).

\subsection{Statistical Analysis}
\label{sect:statistics}

The investigation of clustering properties across the different galaxy samples requires a statistical methodology that can (i) accommodate more than two groups, (ii) remain robust to deviations from normality, and (iii) provide a clear mechanism for pinpointing the specific pairs of samples that drive any observed global differences. To fulfil these requirements, we chose the Kruskal–Wallis H‑test, which tests the null hypothesis that all groups are drawn from the same continuous distribution without assuming any particular shape for that distribution \citep{Kruskal01121952}. In contrast, the two‑sample K-S test, used above to ensure a lack of statistical bias in our pair-matching methodology, is limited to pairwise comparisons and loses power when the number of groups increases, making it unsuitable for a simultaneous assessment of ten samples. By first applying the Kruskal–Wallis test we obtain a single omnibus p‑value that indicates whether at least one sample differs from the others. If this global null hypothesis is rejected, we then employ a Dunn post‑hoc analysis \citep{Dunn01081964} to control for multiple comparisons. The Dunn test computes pairwise rank‑sum differences and yields adjusted p‑values for each sample pair, thereby identifying the specific galaxy populations whose clustering metrics are statistically distinct.

Using this methodology, we observed that the resulting $p$-values of the Kruskal–Wallis H‑test were below the predefined significance threshold ($\alpha = 0.05$) in all instances, leading to the rejection of the null hypothesis and indicating that at least one cluster originated from a distinct distribution. Details of the subsequent post-hoc Dunn’s test in comparison to the BB or GP samples are documented in Table \ref{tbl_post_hoc}. The post-hoc Dunn analysis suggests that the clustering distributions for all evaluated object classes differ significantly from their reference baseline groups (BBs or GPs) across the specified redshift bins, as evidenced by p-values that are nearly all at or approaching zero.

\subsection{Mass of the nearest neighbour}
\label{sect:neighbour}

To further characterise the local environment and its potential impact on galaxy evolution, we investigated the stellar mass of the nearest neighbour for each object in our sample. Environmental factors are critical, as low-mass galaxies often undergo star-formation quenching via rapid gas stripping when situated within the gravitational influence of more massive neighbours (e.g. \citealt{2016MNRAS.463.1916F}).

Figure \ref{fig:hist_nearest} illustrates the mass distribution of the nearest neighbour within a 1 Mpc projected radius for each object in the redshift range $0 < z < 0.1$. Median values for these distributions are denoted by vertical dashed lines and are tabulated in Appendix \ref{tbl_mass}. Our analysis indicates that BBs occupy environments where the nearest neighbour is systematically less massive than those associated with other galaxy populations in the sample.

\section{Discussion}
\label{sect:discussion}

Our observational results, as presented in Figures \ref{fig:clustering_parent_full_range14} and \ref{fig:clustering_parent_full_range23}, show a decrease
in clustering with an increase in SFR (Passive $\rm \rightarrow$ Normal $\rm \rightarrow$ Starburst), which show that star-forming galaxies reside in lower-density environments and passive galaxies are more clustered. Our results are in agreement with previous studies which observed a strong correlation between galaxy clustering and SFR (e.g. \citealt{1980ApJ...236..351D}; \citealt{2005ApJ...630....1Z}; \citealt{2004MNRAS.353..713K}; \citealt{2012MNRAS.419.2670M}), i.e. that red (early-type) galaxies are found in denser environments and are more strongly clustered than blue (late-type, typically star-forming) galaxies, sometimes by a factor of five at small scales. Our findings are in line with the preliminary conclusions of \citet{2009MNRAS.399.1191C}, who observed that GPs do not typically reside in dense environments, although the study relied on a relatively coarse environmental metric and did not explore the issue in depth.

This observed distribution is fundamentally linked to the association of passive galaxies with more massive dark matter halos, as more massive halos inherently cluster more strongly across all cosmic epochs (e.g. \citealt{2005ApJ...630....1Z}). The formation of high-surface-brightness galaxies is statistically biased (e.g. \citealt{1984ApJ...284L...9K}), meaning they are more likely to originate from high-density peaks in the early Universe's matter distribution, leading to their preferential aggregation in dense structures like clusters. In such dense environments, several powerful quenching mechanisms effectively halt star formation: gas within massive halos ($\rm {\sim}  10^{12} M_{\odot}$ or higher) can be heated to high virial temperatures, preventing its efficient cooling and subsequent star formation. AGN, particularly through outflows, inject mechanical energy into the hot gas, suppressing cooling and maintaining the galaxy in a quiescent state (\citealt{2006MNRAS.365...11C}; \citealt{2005MNRAS.362...25B}), and satellite galaxies that fall into the gravitational potential of larger cluster halos can have their cold gas removed by external processes like ram pressure stripping (e.g. \citealt{1972ApJ...176....1G}), strangulation (e.g. \citealt{1980ApJ...237..692L}), and starvation (e.g. \citealt{2000ApJ...540..113B}), thereby ceasing star formation. This overall picture is consistent with the downsizing phenomenon (e.g. \citealt{1996AJ....112..839C}), where more massive galaxies formed the majority of their stars and quenched earlier, aligning with their current prevalence in denser cosmic structures.

In contrast, star-forming galaxies, particularly those with high SFR, are not present in dense clusters and groups, and are found more frequently in voids \citep{2002MNRAS.334..673L}. The tendency of star-forming galaxies to reside in less dense regions is partly due to their role as central galaxies in smaller dark matter halos, which leads to their spatial isolation and results in reduced clustering on small scales \citep{2008ApJ...676..248Y}. In addition, voids lack the efficient quenching mechanisms discussed above that prevent star formation. Therefore, the distribution of galaxies by type reflects a fundamental connection between their internal properties, their evolutionary stage, and the large-scale environments they inhabit \citep{2009ARA&A..47..159B}.

Our results for GPs and BBs, which have low mass ($\rm M\sim 10^{8.5}-10^{10}M_{\odot}$ and $\rm M\sim 10^{6.5}-10^{7.5}M_{\odot}$ respectively) and very high sSFR ($\rm log(sSFR/yr^{-1}) \ge -9.5$), agree with previous observations that field galaxies and in particular dwarf galaxies tend to have a higher sSFR than cluster galaxies \citep{2012ApJ...757...85G, 2025ApJ...978...67M, 2025MNRAS.536..905G}, as well as prior simulations with similar results \citealt{2024ApJ...971..111R, 2025ApJ...978...67M}, and GPs and BBs continue the clustering trend with an even lower number of neighbours, which is expected given that they show on average the highest sSFR values among the compared galaxy samples. We also find that the nearest neighbours to BBs are typically less massive, with a median mass of $\rm log(M_{\star}/M_{\odot}) < 9.1$, than those of other galaxy types where the median mass of the neighbour is $\rm log(M_{\star}/M_{\odot}) > 9.5$, as seen in Figure \ref{fig:hist_nearest}. 
This aligns with findings that galaxies with $\rm log(M_{\star}/M_{\odot}) \sim 8$-$9$ tend to have their star formation quenched when in the presence of large neighbours due to rapid gas stripping \citep{2016MNRAS.463.1916F}.

While GPs and BBs can undergo mergers or interaction events (e.g. \citealt{2022ApJ...933L..11P, 2024ApJ...977...68P} for GPs and \citealt{2017ApJ...846...74P, 2020ApJ...891L..23Z, 2018arXiv180610149R} for BBs), the results from our study suggest these events are very rare for these galaxies, and therefore their sSFR is not commonly driven by mergers.

Another possible driver could be unquenched starburst cycles, which is a scenario in galaxy evolution wherein periods of intense star formation are not effectively shut down or "quenched" by the mechanisms that typically regulate galaxy growth \citep{2009ApJ...692.1305L, 2015MNRAS.446..299M}. However, strong starburst events cannot be sustained for long periods of time and are therefore expected to be transient (e.g. \citealt{1998ARA&A..36..189K}). The lower metallicities of GPs and BBs \citep[][see also Figure \ref{fig:hist_metallicty_ranges}]{2009MNRAS.399.1191C, 2017ApJ...847...38Y, 2024A&A...688A.159K} indicate an absence of substantially long star formation cycles of moderate intensity in the past which would have otherwise increased their metallicities. Figure \ref{fig:hist_metallicty_ranges} also shows that BBs have substantially lower metallicity values than GPs, which may indicate even more recent acquisition of pristine material. A scenario where a dwarf galaxy had low levels of star formation for most of its lifetime, and only recent strong starbursts, seems the most likely to explain the low metallicity found in these GPs and BBs.

Most high redshift galaxies occupy a similar or even lower metallicity range and display the compact morphology characteristic of emission-line-dominated systems, thereby placing GPs as their low-redshift counterparts \citep{2025arXiv250522600C}. There are 19 LRD analogues (\citealt{2022A&A...665L...4S}; \citealt{2023ApJ...942L..14R}; \citealt {2025ApJ...980L..34L}; \citealt{2025arXiv250710659L} and \citealt{2025arXiv251002801C}), of which 7 are GPs, suggesting that the large-scale environments of these LRDs could share some characteristics with the GPs and BBs in our study. One could envisage a similar formation scenario for the GPs and BBs, since they also reside in low density regions away from the halos of more massive galaxies. Moreover, \cite{2025MNRAS.539.2910P} suggest that LRDs are too abundant to be hosted by the same dark matter halos as unobscured quasars as would be expected from the formation scenario.

Overall, the GPs and BBs only represent a small minority of dwarf galaxies observed in our sample, while simultaneously being the most actively star forming. This is consistent with their proposed position as rare and extreme transient events occurring in isolated environments where pristine gas is available for recent starbursts. A case-by-case analysis of each GP and BB would be necessary to investigate which mechanism is responsible for their ongoing starburst. In particular, finding the traces of a past merger responsible for the ongoing starburst or looking for signs of recurrent starburst cycles would prove challenging, and would require higher resolution data to probe e.g. star formation history, metallicity gradients, morphology, or stellar populations within each GP or BB.

To assess whether simulations can help interpret our results, we searched for GP and BB analogues in the large cosmological simulation TNG300 from the Illustris The Next Generation (TNG) suite \citep{2018MNRAS.473.4077P, 2019ComAC...6....2N, 2020ApJS..248...32W}, performed with the AREPO code \citep{2010MNRAS.401..791S}. Applying the photometric selection criteria of \cite{2017ApJ...847...38Y} and \cite{2009MNRAS.399.1191C} to identify BB and GP galaxies at $z=0.03$ and $z=0.3$ respectively, we found no matches in TNG300. However, this absence does not necessarily imply that such analogues are lacking, as IllustrisTNG galaxies are known to be systematically redder than observations, particularly at low masses \citep{2018MNRAS.475..624N}, which may render these criteria unsuitable for simulations. As an alternative, we identified candidate BB and GP analogues based on their location in the stellar mass–sSFR plane (Fig.~\ref{fig:scatter_plot_redshift}), combined with a compactness criterion (half-mass radius $R_\mathrm{h}<1$~kpc). However, the vast majority of candidates selected in this way, while exhibiting physical properties similar to observed GPs and BBs (i.e. compact, starbursting dwarfs), were found to be typically orbiting massive host galaxies. Their elevated SFRs were therefore likely driven by environmental effects such as mergers or flybys, in contrast with observed BBs and GPs, which are preferentially found in low-density environments. We thus concluded that isolated, compact, starbursting dwarf galaxies analogous to observed BBs and GPs are not well reproduced in Illustris. We defer a detailed investigation of this discrepancy to a future paper, as it may point to limitations in current cosmological simulations in forming such systems.


\section{Conclusion}
\label{sect:conclude}

In this study, we investigated the large-scale environments of Green Pea (GP) and Blueberry (BB) galaxies by analysing their clustering properties relative to a control sample from SDSS, matched in stellar mass and specific star-formation rate. Our analysis uses the number of neighbours within 5 Mpc as a proxy for environmental density and confirms a robust anti-correlation between sSFR and clustering strength across all stellar masses. The results can be summarised as follows:

\begin{itemize}
  
  \item[\color{black} $\bullet$]  We find that GPs and BBs, distinguished by their exceptionally high sSFR, represent the extreme end of the trend where galaxies with higher sSFR tend to have fewer neighbours, exhibiting the weakest clustering in our sample. The most extreme are BBs, for which we also observe that the nearest neighbours are less massive than those of other galaxy types. This result firmly establishes that these galaxies predominantly reside in low-density environments.


  \item[\color{black} $\bullet$] The presence of GPs and BBs in isolated environments indicates their intense starburst activity is not primarily driven by environmental mechanisms common in dense regions, such as frequent major mergers or starburst cycles. Instead,
  the combination of their low-density environment and low metallicity suggests these are relatively un-evolved systems undergoing a recent, transient, and powerful starburst after a largely quiescent history. This favours internal processes, such as star formation cycles or the accretion of pristine gas from the intergalactic medium, as the more likely triggers for their extreme activity. 

  \item[\color{black} $\bullet$] These findings strengthen the role of GPs and BBs as essential local analogues for the high-redshift, low-mass galaxies believed to have contributed to cosmic reionization, which are also theorized to form in less dense regions of the early Universe.

  
\end{itemize}

\begin{acknowledgements} MG and JS thank GACR project 22-22643S for the support and the institutional support from the Astronomical Institute of the Czech Academy of Sciences, code RVO:67985815.\\
KK is supported by the INTER-COST LUC24023 project of the INTER-EXCELLENCE II programme, funded by the Czech Ministry of Education, Youth and Sports.\\
NP is supported by the European Research Council (ERC) under grant agreement no.\ 101040751.\\

MG thanks Matheen Musaddiq for invaluable technical expertise.\\

The SDSS is managed by the Astrophysical Research Consortium for the Participating Institutions. The Participating Institutions are the American Museum of Natural History, Astrophysical Institute Potsdam, University of Basel, University of Cambridge, Case Western Reserve University, University of Chicago, Drexel University, Fermilab, the Institute for Advanced Study, the Japan Participation Group, Johns Hopkins University, the Joint Institute for Nuclear Astrophysics, the Kavli Institute for Particle Astrophysics and Cosmology, the Korean Scientist Group, the Chinese Academy of Sciences (LAMOST), Los Alamos National Laboratory, the Max-Planck-Institute for Astronomy (MPIA), the Max-Planck-Institute for Astrophysics (MPA), New Mexico State University, Ohio State University, University of Pittsburgh, University of Portsmouth, Princeton University, the United States Naval Observatory, and the University of Washington.\\

The authors would like to express their gratitude to the anonymous reviewer for their insightful comments and suggestions, which have significantly improved the quality of this manuscript.

\end{acknowledgements}

\bibliographystyle{aa}
\bibliography{sample63}

\begin{appendix}
\section{Additional Figures \& Tables}

\begin{figure*}
\centering
\includegraphics[width=0.99\linewidth]{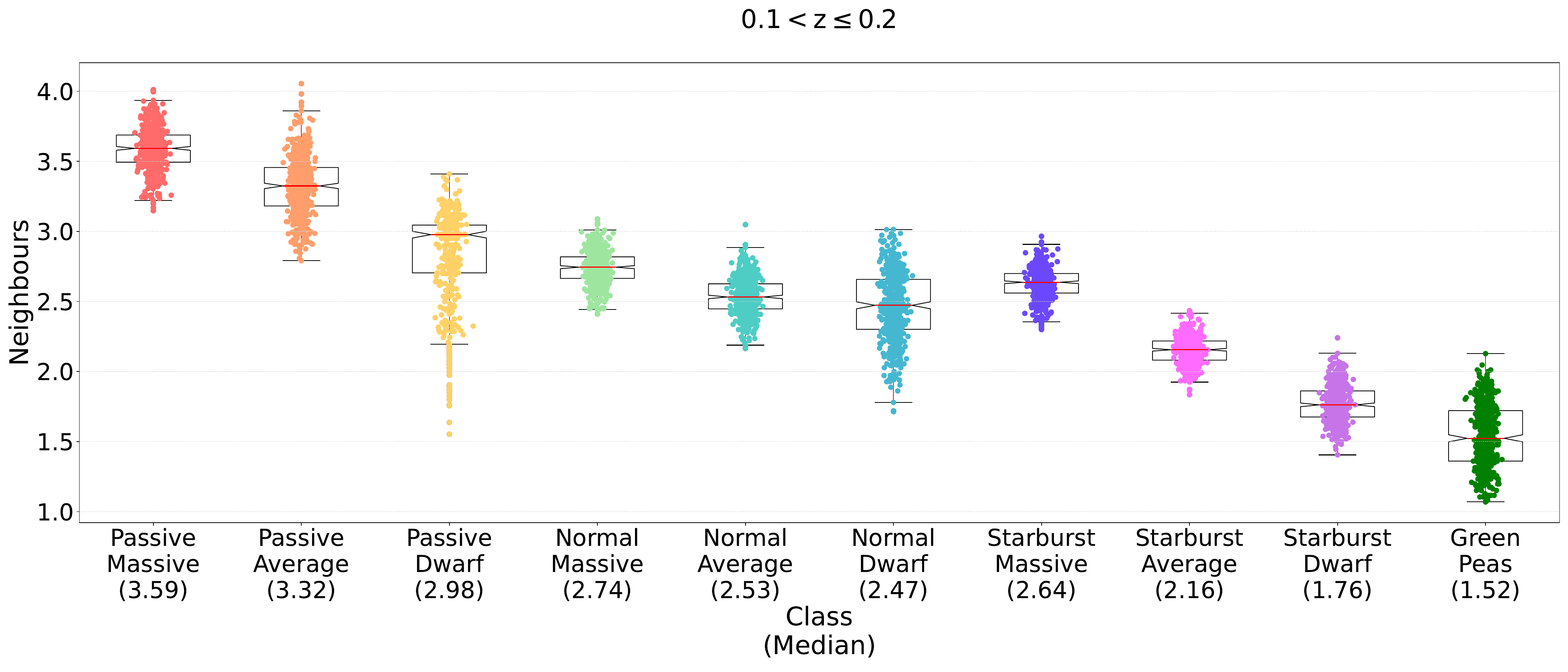}
\includegraphics[width=0.99\linewidth]{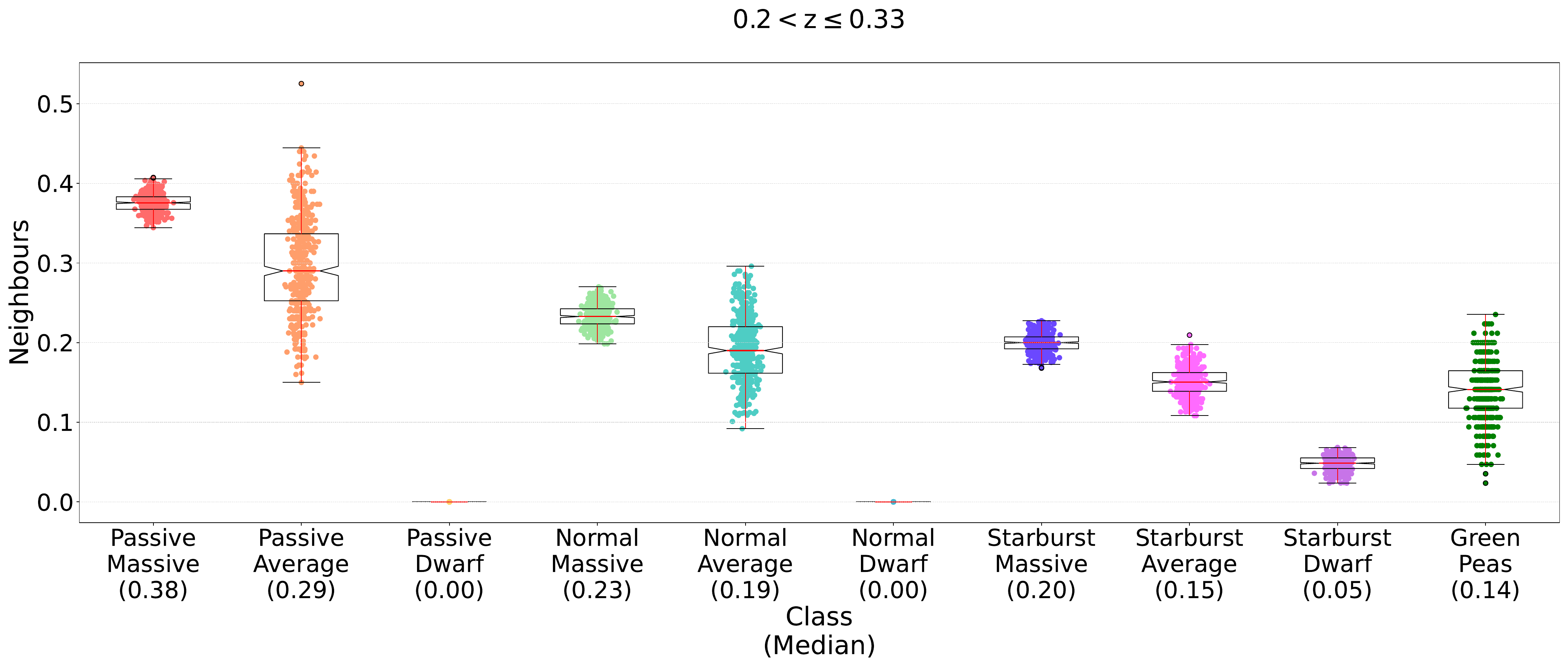}
\caption{Box plot showing the average number of neighbours within 5 Mpc for the 500 bootstrapped samples for each class of object in the redshift range $0.1 < z \le 0.2$ (top panel) and $\rm 0.2 < z \le 0.33$ (bottom panel). Each point is the mean of the number of neighbours for a bootstrapped sample, resulting in 500 points per class. The median number of neighbours for each class is shown with a red line, and is recorded numerically below the class names. Due to the very small size of the Passive Dwarf and Normal Dwarf samples in the redshift range $\rm 0.2 < z \le 0.33$, their statistics are poor.}
\label{fig:clustering_parent_full_range23}
\end{figure*}

\begin{table}[h!]
\caption{Metallicity of objects in
redshift ranges, noted in Figure  \ref{fig:hist_metallicty_ranges}}
\label{tbl_metal}
\begin{center}
 \begin{tabular}{c c c c} 
 \hline
 Object Class &  $0 < z \le 0.1$ & $0.1 < z \le 0.2$ & $0.2 < z \le 0.33$ \\ [0.5ex] 
 \hline\hline
Passive Massive & 9.12 & 9.08 & 9.09 \\
Passive Average & 8.95 & 9.00 & 8.96\\
Passive Dwarf & 8.66 & 8.70 & 8.53 \\
Normal Massive & 9.08 & 8.96 & 9.00 \\
Normal Average & 8.91 & 8.92 & -\\
Normal Dwarf & 8.68 & - & - \\
Starburst Massive & 9.19 & 9.11 & 9.16 \\
Starburst Average & 8.83 & 8.85 & 8.94 \\
Starburst Dwarf & 8.57 & 8.61 & 8.66 \\
BBs & 7.78 & - & - \\
GP Sample 1 & - & 8.59 & - \\
GP Sample 2 & - & - & 8.70 \\
 \hline
\end{tabular}
\end{center}
\end{table}

\begin{table}[h!]
\caption{Median of the histogram of the mass of nearest neighbour of objects in redshift range $0 < z \le 0.1$, noted in Figure  \ref{fig:hist_nearest}. }
\label{tbl_mass}
\begin{center}
 \begin{tabular}{c c} 
 \hline
 Object Class & Mass ($\rm log(M_{\star}/M_{\odot}) $) \\ [0.5ex] 
 \hline\hline
Passive Massive & 9.99 \\
Passive Average & 10.03 \\
Passive Dwarf &  9.81 \\
Normal Massive &  9.99 \\
Normal Average &  9.95 \\
Normal Dwarf & 9.72 \\
Starburst Massive &  10.04 \\
Starburst Average & 9.80 \\
Starburst Dwarf & 9.63 \\
BBs & 9.12 \\
 \hline
\end{tabular}
\end{center}
\end{table}

\end{appendix}
\end{document}